\pdfoutput=1

\documentclass{appolb}

\usepackage{cite}
\usepackage{wrapfig}
\usepackage{graphicx}
\usepackage{amssymb}
\usepackage{amsfonts}
\usepackage{amsmath}
\usepackage{longtable}
\usepackage{rotating}
\usepackage{lscape}
\usepackage{epsfig}
\usepackage{multirow}

\usepackage{xcolor}

\usepackage{lineno}

\usepackage[percent]{overpic}

\usepackage{hyperref}
\usepackage[all]{hypcap}
\hypersetup{
    colorlinks,
    citecolor=blue,
    filecolor=blue,
    linkcolor=black,
    urlcolor=blue
}

\def\pt{p_{\rm T}}

\def\av#1{\langle #1 \rangle}

\begin{document}
\title{ Cumulants of net-charge distribution 
from particle-antiparticle sources%
\thanks{Presented at Excited QCD 2020.}%
}
\author{Igor Altsybeev
\address{ Saint-Petersburg State University \\ Universitetskaya nab. 7/9, St. Petersburg, 199034, Russia  }
\\
\textit{i.altsybeev@spbu.ru}
}
\maketitle
\begin{abstract}

It is shown
how  high-order cumulants of net-charge distribution in hadronic collisions at LHC energies
can be 
expressed via lower-order terms
 under the assumption 
that particle-antiparticle pairs are  produced in 
independent local processes.
It is argued and tested with HIJING model
 that this assumption is typically valid for net-proton fluctuations 
in case when no critical behaviour is present in the system.
Values estimated in such a way can be considered 
as baselines for direct measurements of high-order net-charge fluctuations in real data.

\end{abstract}

  
\section{Introduction}

In heavy-ion collision experiments, 
measurements of high-order 
fluctuations of conserved quantities, 
such as net-charge, net-baryon, net-strangeness, 
are of great importance 
since 
they should increase 
in the vicinity of the critical point of the QCD diagram \cite{Shuryak_et_al_1999}
and may serve as a
signature of the transition between the hadronic and partonic phases.
These expectations are confirmed also by lattice QCD calculations  \cite{bazavov_2012}.
At LHC energies, 
 a smooth crossover  between a hadron gas and the QGP is expected
 \cite{bazavov_2012, Borsanyi:2018grb}.
Studies of the net-particle cumulant ratios 
are of a 
special interest 
because of their direct  connection to 
susceptibilities theoretically calculable 
in the lattice QCD.
In particular, 
net-proton fluctuations 
 have been extensively studied experimentally  \cite{Luo:2015ewa, ALICE_netproton}.



The net-charge is defined as  $\Delta N = N^+ - N^-$,
where  $N^+$ and $N^-$ are
the numbers of positively and negatively charged particles
measured  in an event within rapidity  acceptance $Y$ (i.e. $y\in -Y/2, Y/2$).
For example, the second cumulant of net-charge distribution is given  by 
\begin{equation}
\kappa_2 (\Delta N) =  \big< \big(\Delta N- \av{\Delta N}\big)^2 \big> = \av{(\Delta N)^2}- \av{\Delta N}^2,
\end{equation}
where angular brackets denote averaging over events.
Expressions for higher-order cumulants 
are increasingly more complicated.

If fluctuations of both $N^+$ and $N^-$ are Poissonian, 
$\Delta N$ has the 
Skellam distribution, with cumulants
$\kappa_r (\Delta N) = \av{N^+} +  (-1)^r\av{N^-}$,
$r=1,2,...$, such  that, for instance,  ratio $\kappa_4/\kappa_2=1$.
%
The Poissonian particle production
is usually considered as a baseline model.
However, in reality
the cumulants of particle distributions are very sensitive to
 two so-called {\it non-dynamical} contributions
that are not related to criticality in the system.
The first contribution comes from the fluctuations of a number of emitting sources -- the so-called
``volume fluctuations'' (VF) \cite{PBM_AR_JS_NPA, VF_Nonaka_2019}.
The second contribution is 
due to charge conservation laws, for example, 
from neutral resonances decaying into pairs of oppositely charged particles.
These 
two effects make interpretation of  experimental measurements of the cumulants highly non-trivial, especially for higher-order cumulants.
Both of them should be taken into account when one tries to extract signals of critical behaviour from measured observables
\cite{Bzdak_2013_baryon_number_conserv,
PBM_AR_JS_NPA, PBM_AR_JS_2019}. 
At LHC energies, however,
it is possible
to construct a simple baseline model 
that include both these effects,
if one assumes production of oppositely charged pairs
that are nearly uncorrelated in rapidity.
This is demonstrated in Section \ref{sec:decomp}, and 
tested with HIJING event generator in Section \ref{sec:ratios_in_HIJING}
for the case of net-proton fluctuations.
More details can be found in \cite{Altsybeev:2020qnd}.

Yet another caveat about ordinary cumulants $\kappa_r$  is that they
 get trivial contributions to all orders due to self-correlations.
It was shown in  
\cite{Kitazawa:2017ljq, OllitraultIsolatingNetCharge2018}
that
self-correlations can be removed systematically by constructing {\it factorial} cumulants. This is briefly considered in  Section \ref{sec:fc}.

\section{Decomposition of cumulants for two-particle sources}
\label{sec:decomp}


Creation of  oppositely charged particle 
pairs 
is governed by a local charge conservation.
The simplest case of a pair production process is a two-body neutral resonance decay,
where integer $+1$ and $-1$ charges are produced,
and net-charge contribution to cumulants 
from a resonance  is determined solely by its decay kinematics and resonance spectra.
Another process is  string fragmentation that produces fractional charges 
at each  breaking point (quarks, diquarks),
which then combine with  partons from 
next breaking points. 
That may lead to a  correlation 
between hadrons coming from several adjacent parts of a string, and, therefore, influence 
net-charge fluctuations in a complicated way.

In case of protons and antiprotons, however,
there are no resonances that decay into $p$ and $\overline{p}$ pair.
Such $p$-$\overline{p}$ pairs are produced mainly in string breaking
(each of them may be produced directly or via a decay of a short-lived resonance).
Moreover, a probability of production of two or more baryon pairs
from adjacent parts of the same string  is  low.
Registration of  $p$-$\overline{p}$ pairs from jets in a low transverse momentum ($\pt$) range (typically one takes $\pt \lesssim 2$ GeV/$c$)
 should be low as well.
Therefore, if there are no processes  
other than resonance decays 
and string fragmentation,
 the $p$-$\overline{p}$ pairs visible in an event
may be considered as nearly independent.
This allows one to write simplifying expressions 
for the cumulants of net-charge fluctuations as it is described below.

Decompositions
of cumulants 
 for a system of $N_S$ independent sources 
up to the fourth order are provided in \cite{PBM_AR_JS_NPA}
and up to the eighth order -- in \cite{Altsybeev:2020qnd}.
At LHC energies, where $\av{N^+}=\av{N^-}$,
 the second and the fourth cumulants of net-charge distribution
decompose as
\begin{equation}
\kappa_2(\Delta N) = k_2 (\Delta n )  \av{N_S},
\end{equation}
\begin{equation}
\kappa_4(\Delta N) = k_4 (\Delta n) \av{N_S}
 + 3 k_2^2(\Delta n) K_2(N_S),
\end{equation}
where $\Delta n=n^+ - n^-$ is a net-charge of a single source,
and different notations for cumulants $\kappa$, $k$ and $K$
 serve only
for better visual distinction 
which distribution they are referred to.
The ratio of the fourth to the second cumulant 
reads as
\begin{equation}
\label{k4_to_k2_indep_emitters}
\frac{\kappa_4}{\kappa_2 }(\Delta N) = \frac{k_4}{k_2} (\Delta n)
 + 3 k_2(\Delta n) \frac{K_2(N_S)}{\av{N_S}}.
\end{equation}
The VF enter this equation via the second term that is proportional to the variance of the number of sources.
The formulae above are valid for any types of sources,
in particular, in \cite{PBM_AR_JS_NPA} ``wounded nucleons'' are considered. 
Instead, we may treat the sources 
as {\it particle-antiparticle pairs},
for instance, $p$-$\overline{p}$. 
Note that these sources may be correlated
to a certain extent (for example, due to radial and azimuthal flow)
provided that swapping of the charges in each produced pair 
does not affect the physics of the whole event. 
The fourth cumulant 
for a single two-particle source
simplifies to
\begin{equation}
\label{k4_2part_source}
k_4(\Delta n) =k_2(\Delta n)  - 3k_2^2(\Delta n) .
\end{equation}
We may now recall the argument that  $p$-$\overline{p}$ pairs are nearly 
uncorrelated in rapidity 
and the fact that the distribution of $p$$(\overline{p})$ is nearly flat at mid-rapidity 
$|y| \lesssim 1$ at LHC energies. It turns out that in this case
it is possible to rewrite the cumulant ratio \eqref{k4_to_k2_indep_emitters} 
using    quantities that are measurable in an experiment \cite{Altsybeev:2020qnd}:
\begin{equation}
\label{k4_to_k2_VIA_R2_Np}
	\frac{\kappa_4}{\kappa_2 }(\Delta N) 
	= 1 + 3 \kappa_2(\Delta N) R_2(N^+), 
\end{equation}
where $R_2(N^+) =   \av{N^+(N^+-1)}  / \av{N^+}^2  - 1$
is the so-called robust variance of a number of positive particles measured within 
acceptance $Y$ (equivalently, $R_2(N^-)$ could be used instead).
Values of the cumulant ratio  calculated with \eqref{k4_to_k2_VIA_R2_Np}
could be considered as baselines for experimental measurements of the ratios (instead of, for instance, the Skellam baseline). 
Possible signals from critical phenomena would be indicated by 
some deviations from these baselines.

\section{Application to realistic model}
\label{sec:ratios_in_HIJING}

\begin{figure*}[b] 
\centering 
\begin{overpic}[width=0.49\textwidth, trim={0.3cm 0.0cm 0.5cm 0.5cm},clip] 
{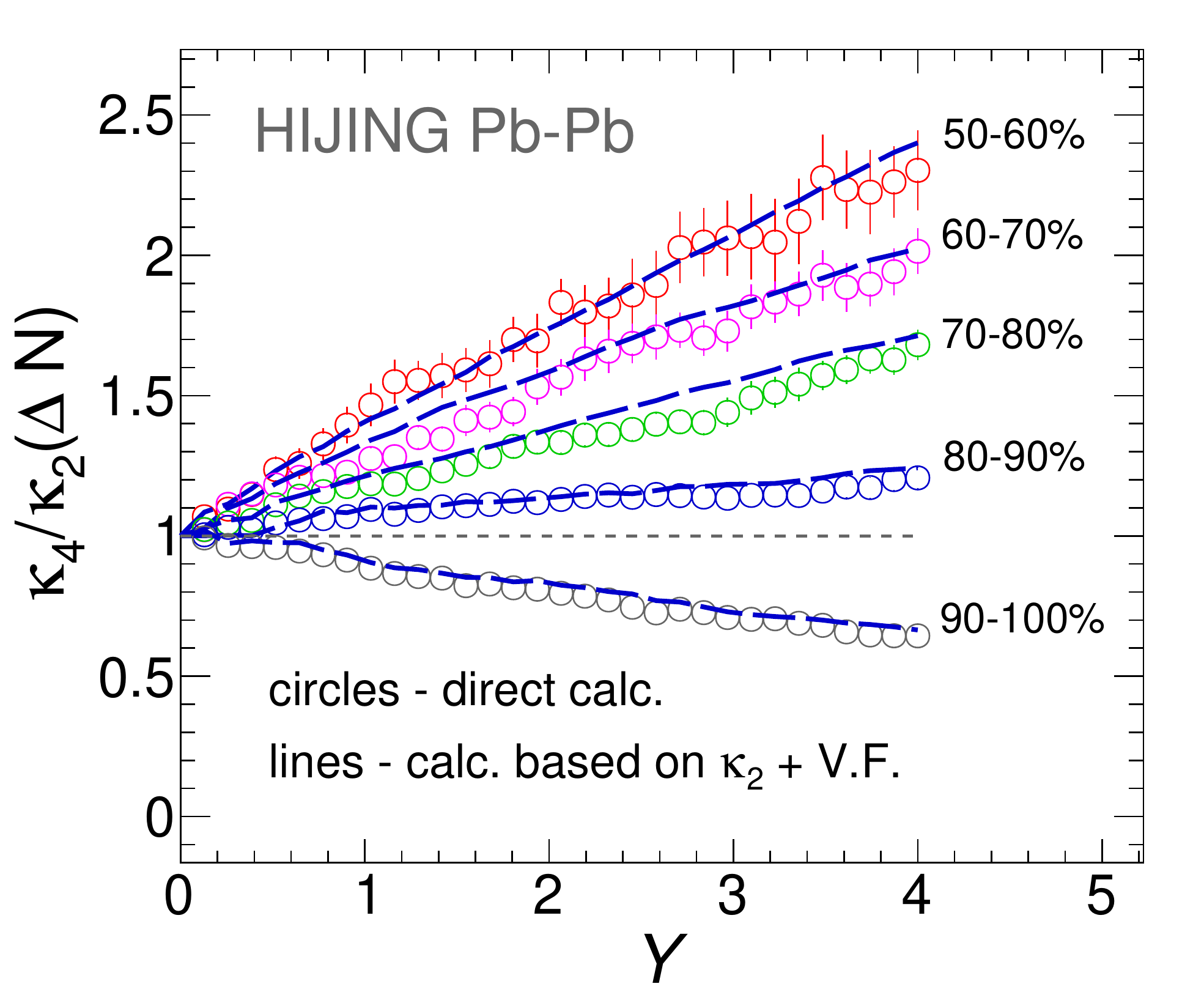} 
\put(20.4,67){\footnotesize \color{darkgray} $p_{\rm T}$$\in$0.6--2 GeV/$c$ }
\put(20.4,61){\footnotesize \color{darkgray} net-proton}
\put(65,76.6){\footnotesize centrality:}
\end{overpic}
\begin{overpic}[width=0.49\textwidth, trim={0.3cm 0.0cm 0.5cm 0.5cm},clip]
{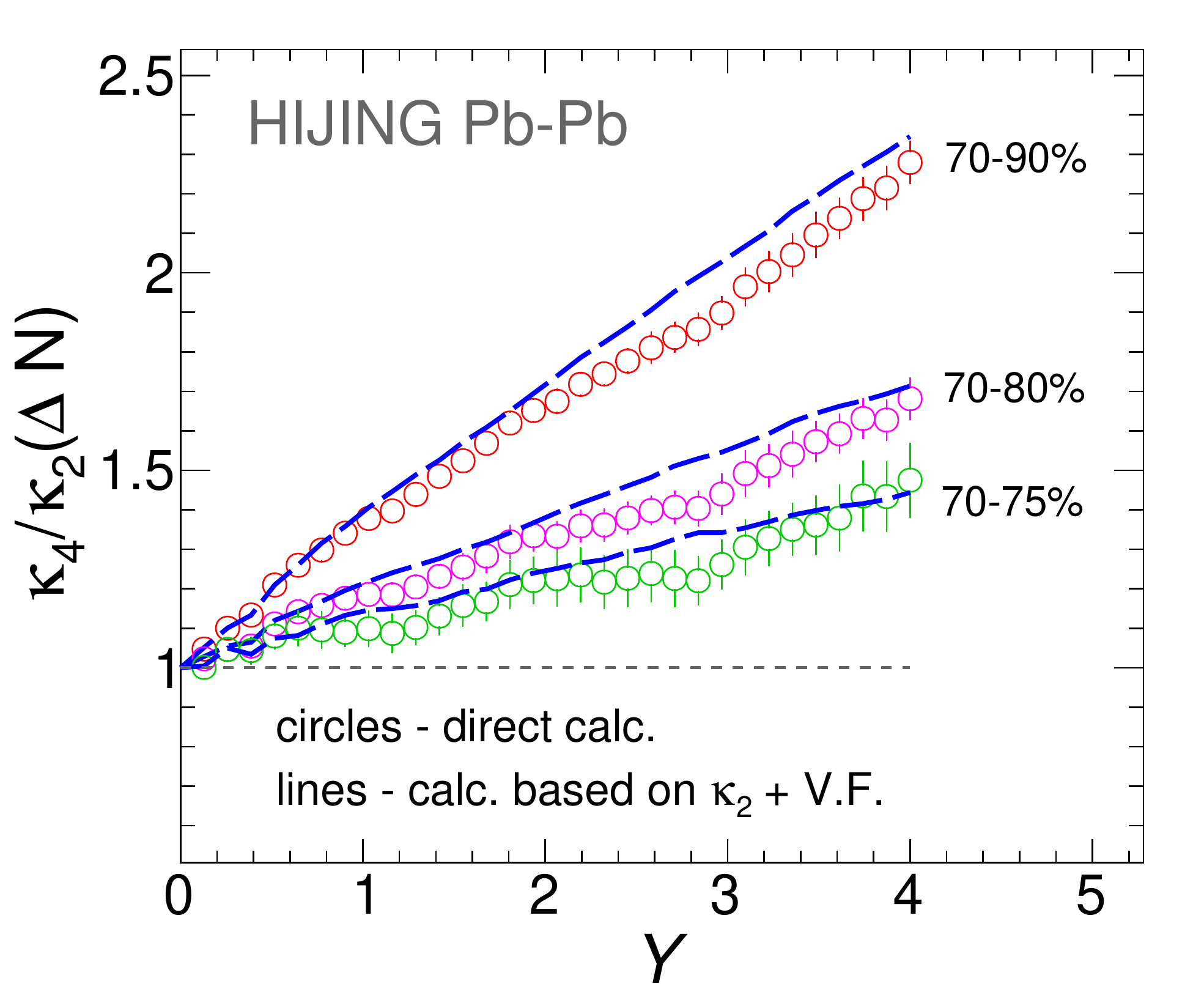} 
\put(20.4,67){\footnotesize \color{darkgray} $p_{\rm T}$$\in$0.6--2 GeV/$c$ }
\put(20.4,61){\footnotesize \color{darkgray} net-proton}
\put(65,76){\footnotesize centrality:}
\end{overpic}
\caption{ 
Dependence 
 of the 
net-proton $\kappa_4/\kappa_2$  ratio  
on the size of the rapidity acceptance $Y$
in HIJING in Pb-Pb events 
at $\sqrt{s_{NN}}=2.76$ TeV \cite{Altsybeev:2020qnd}. 
Direct calculations are shown by circles,
analytical calculations with \eqref{k4_to_k2_VIA_R2_Np}
-- by dashed lines.
Panel (a) -- results in several centrality classes 
of the class width 10\% 
 are shown, 
(b) -- dependence on the width of centrality class (20\%, 10\% and 5\%)
is demonstrated.
Note that in each graph  there are point-by-point correlations
as $Y$ increases.
}
\label{fig:ratios_in_HIJING} 
\end{figure*}
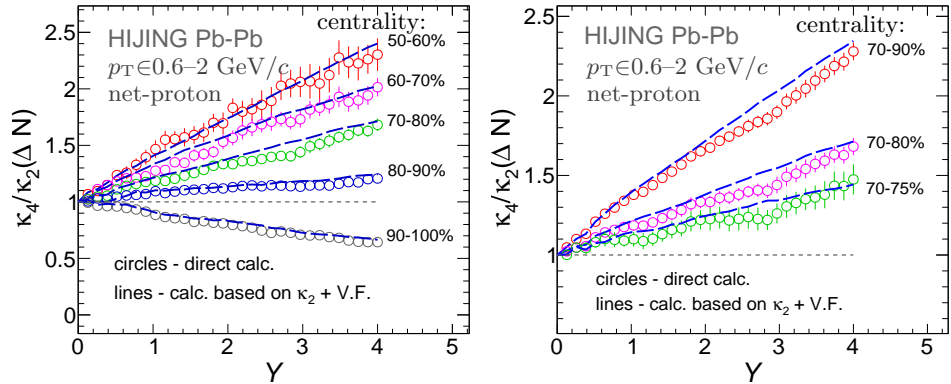

Validity of the assumptions about charged pair production
done above was  put into test 
using HIJING monte-carlo generator, 
which simulates  multiple jet production and fragmentation of quark-gluon strings 
\cite{Wang:1991hta}.
For that, 
analysis of net-proton fluctuations
in Pb-Pb collisions simulated in HIJING was performed \cite{Altsybeev:2020qnd}. 
Protons and antiprotons within $y\in(-2, 2)$
and transverse momentum 
range 0.6--2 GeV/$c$
were selected.
Figure \ref{fig:ratios_in_HIJING} (a) shows the
dependence on rapidity acceptance   $Y$
of the  $\kappa_4/\kappa_2$   ratios
 calculated directly (circles)
and by
expression  \eqref{k4_to_k2_VIA_R2_Np} (lines)
 in several centrality classes. 
Centrality was  determined using multiplicity distribution in two symmetric
$3<|\eta|<5$ ranges, which emulates the way
of centrality  determination in real experiments.
A good agreement between the calculations
can be seen in all classes, 
indicating that 
the assumption about the $p$-$\overline{p}$
 pairs as nearly independent sources 
is approximately valid in HIJING.
Slopes of the lines for different centrality classes reflect changes in VF
via the second term in \eqref{k4_to_k2_VIA_R2_Np}.
Panel (b) demonstrates a decrease of   $\kappa_4/\kappa_2$ 
values with the width of a centrality class (when the width
changes from  20\%  down to 5\%), 
which is explained by a reduction of the volume fluctuations with the narrowing of the class.
It was checked also that 
the robust variance $R_2(N^+)$   
as a function of  
$Y$
stays constant,
which is  essential for calculations  with~\eqref{k4_to_k2_VIA_R2_Np}.
More details of this study, in particular, a decomposition expression
for the  $\kappa_6/\kappa_2 (\Delta N)$ ratio
 can be found in \cite{Altsybeev:2020qnd}.

\begin{figure*}[b] 
\centering 
\begin{overpic}[width=0.6\textwidth, trim={0.1cm 0.08cm 0.7cm 0.7cm},clip] 
{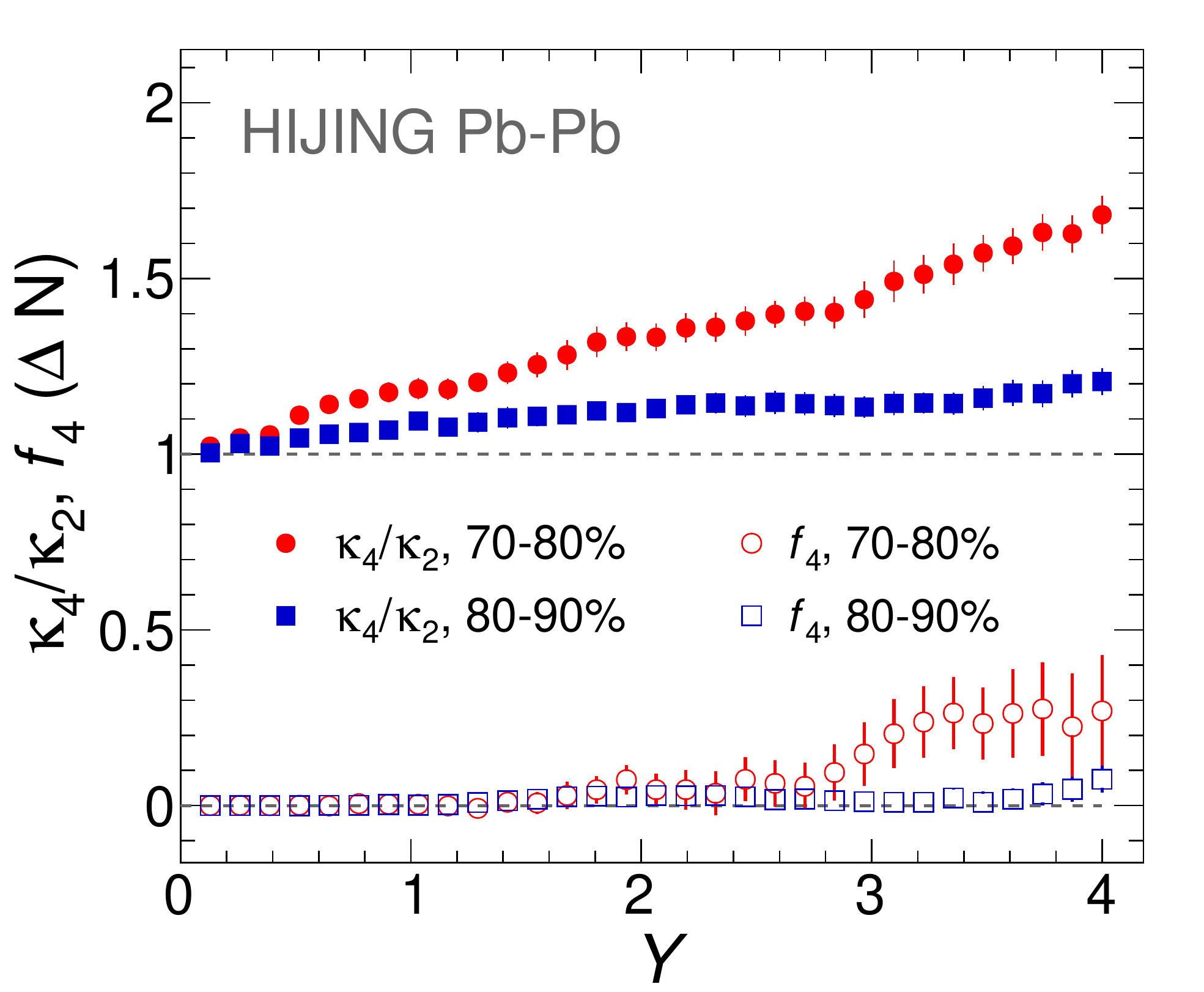} 
\put(20.4,66.5){\footnotesize \color{darkgray} $p_{\rm T}$$\in$0.6--2 GeV/$c$ }
\put(20.4,61){\footnotesize \color{darkgray} net-proton}
\end{overpic}
\caption{ 
Dependence net-proton cumulant ratio $\kappa_4/\kappa_2$   (closed markers) and factorial cumulant $f_4$
(open markers)
on $Y$ in HIJING. Pb-Pb collisions
at $\sqrt{s_{NN}}=2.76$ TeV,
centrality classes  70-80\% and 80-90\%.
}
\label{fig:k4k2_f4_in_HIJING} 
\end{figure*}

\section{Factorial cumulants}
\label{sec:fc}
\noindent The fourth-order factorial cumulant of net-charge distribution is given by
\begin{multline}
\label{eq:f4}
f_4   =  \kappa_4   
-6 \big( \av{NQ^2} - \av{N}\av{Q^2} - 2 \av{NQ}\av{Q}+2\av{N}\av{Q}^2 \big) \\
+ 8\big(\av{Q^2}-\av{Q}^2\big) 
+3\big(\av{N^2} - \av{N}^2\big)
- 6\av{N} ,
\end{multline}
where $N = N^+ + N^-$ and $\Delta N$ is denoted as $Q$ for clarity \cite{OllitraultIsolatingNetCharge2018}.
It is interesting to check the behaviour of this observable 
in realistic models. 
As an example, Pb-Pb collisions from
HIJING  were analyzed in the present work.
Values of the factorial cumulant $f_4$ 
 of net-proton distribution
and the conventional $\kappa_4/\kappa_2$ ratios  
are shown in Figure \ref{fig:k4k2_f4_in_HIJING} as a function of the acceptance width $Y$
in two centrality classes 70-80\% and 80-90\%\footnote{Results are shown for 
two peripheral centrality classes only, since 
for more central classes statistical uncertainties for $f_4$ are much larger.
}.
The $\kappa_4/\kappa_2$ ratios 
are the same as in Fig. \ref{fig:ratios_in_HIJING},
and the values
are above unity (i.e. above the Skellam baseline)
due to the VF,
as it was discussed above. 
Moreover, values in class 70-80\% are higher than in 80-90\%
since the  VF in the former class are larger.
In contrast, factorial cumulants
 $f_4$  are compatible with zero for both centralities.
This is because factorial cumulants 
of order $k$ remove contributions
 of lower orders $r<k$,
which 
means, in particular, 
that net-proton  $f_4$ 
should be suppressed in HIJING.
Factorial cumulants are also much less sensitive to the VF than ordinary cumulants
\cite{OllitraultIsolatingNetCharge2018}.
However, in distinction from ordinary cumulants, 
factorial cumulants  cannot be directly compared with the lattice data \cite{Kitazawa:2017ljq, OllitraultIsolatingNetCharge2018},
therefore their usefulness in studies of the QCD diagram is under question.

\section{Summary}

It   was shown that 
 high-order cumulants of net-charge distribution 
can be decomposed into lower-order terms
under the assumption of
independent production of particle-antiparticle pairs.
At LHC energies,
this should be a good approximation for net-proton fluctuations at mid-rapidity
in case if there is no critical behaviour in the system,
as it
 was demonstrated for the $\kappa_4/\kappa_2$ ratio in HIJING.
Such reduced expressions for high-order cumulants
can be considered as baselines for direct experimental measurements.
 It was shown also  that the fourth net-proton factorial cumulant  in HIJING is compatible 
with zero, which also indicates that there are no sources
of genuine high-order net-proton correlations in this generator.

\section*{Acknowledgements}
This work is supported by the Russian Science Foundation, grant 17-72-20045.

\bibliographystyle{h-elsevier}

\bibliography{bibliography}

\end{document}